# Mesoscale Modelling of Confined Split-Hopkinson Pressure Bar Tests on Concrete: Effects of Internal Damage and Strain Rates

Qingchen Liu and Yixiang Gan [*]

School of Civil Engineering, The University of Sydney, NSW 2006, Australia

[*] Corresponding author: yixiang.gan@sydney.edu.au

**Abstract:** The dynamic strength of concrete under complex loading conditions is a key consideration in the design and maintenance of infrastructures. To assess this mechanical property, Split Hopkinson Pressure Bar (SHPB) tests are typically adopted across a wide range of loading and confining conditions. In this study, mesoscale modelling based on the finite element method (FEM) is employed to simulate SHPB tests on three-phase concrete with realistic aggregate shape, in order to investigate the effects of loading ramp rate, internal friction, and confining pressure on the dynamic increase factor (DIF). Microscopic evidence to explain these effects is explored through analysing the distributions of the internal strain rate and local damage. As key results, increasing loading ramp rates, internal friction, and confining pressure can generally leads to higher DIF values. Only a higher loading ramp rate significantly amplifies the strain-rate effect on the DIF, as evidenced by pronounced increases in both internal strain rate and damage in the mortar and aggregate phases. In contrast, higher internal friction and confining pressure weaken the strain-rate effect on the DIF. Both can be attributed to the mortar phase, which shows a less pronounced increase in damage with increasing strain rate. This study enriches the understanding of the dynamic fracture of concrete toward complex loading scenarios.

# Highlights:

- Mesoscale modelling of confined dynamic tests on concrete is conducted and validated.
- For SHPB tests, a higher loading ramp rate enhances the strain rate effect on concrete strength.
- Under confining pressure, the dynamics strength is increased due to extensive damage in aggregates.
- Local damage and strain rate information is used to explain these findings.

# 1. Introduction

As the most widely used construction material worldwide, concrete plays an essential role across a wide range of onshore and offshore infrastructure applications, such as bridges [1], tunnels [2], concrete dams [3], and offshore wind turbines [4]. Throughout their service life, concrete structures are exposed to extreme weather and complex geotechnical conditions and may suffer natural or accidental events, such as earthquakes [5, 6], rockfall events [7, 8], tsunamis [9, 10] and ship impacts [11, 12], leading to potentially huge and irreversible damage to concrete structures. Thus, characterisation of concrete behaviour under dynamic loading is a prerequisite for safe and efficient structural design [13-15]. Various types of dynamic loading, e.g., impact [16], blast [17], seismic [18], and cyclic loadings [19], have been considered. As applied stress conditions become more complex to reflect practical scenarios, it becomes progressively more challenging to precisely control over the loading conditions for investigating the dynamic response of concrete to external conditions. In addition to the vastly existed unconfined experimental cases, confined dynamic loading has also received increasing attention [20, 21]. As a typical heterogeneous material, concrete has a microstructure spanning length scales from the nano- to the macroscale [22-24], and its complicated internal structure of concrete poses challenges in identifying the underlying mechanisms of dynamic fractures. The mesoscale allows concrete to be treated as a three-phase composite consisting of mortar, aggregates, and the interfacial transition zone (ITZ), making it possible to represent the main material heterogeneity of concrete and to capture its overall and local responses [25]. Yet, due to the intrinsic complexities in the material phases and temporospatial variations of the strain rates, there is still a lack of investigations on bridging microscopic information to the macroscopic findings under dynamic stress conditions.

The dynamic response of concrete over a wide range of strain rate has been investigated using various approaches, such as drop hammer tests [26], Split Hopkinson Pressure Bar (SHPB) tests [27], bullet impact tests [28], and explosion tests [29]. For the broadly focused strain rate range relevant to the infrastructure applications (around $10^{-5} \sim 10^3$ s$^{-1}$), the SHPB test is the most widely regarded approach and has been used to characterise the overall response of concrete at strain rates of about $10 \sim 10^3$ s$^{-1}$ [27], whereas drop hammer tests are typically used for lower strain rates up to around $10$ s$^{-1}$ [26]. As commonly observed, with increasing strain rate, the stress-strain behaviour of concrete becomes more ductile with its instantaneous material properties changing accordingly. The most notable effect is the significant increase in concrete strength [30], quantified by the dynamic increase factor (DIF) as the ratio of dynamic to static strength [31]. Based on the available experimental data, many attempts have been made to capture the relationship between DIF and strain rate through empirical models, such as those proposed by the International Federation for Structural Concrete (FIB) [32], Bischoff and Perry [33], and Ross and Tedesco [27]. The strain rate dependency of concrete strength can be divided into low and high strain rate regions, characterised by the low and high rate sensitivity, respectively, and separated by a critical strain rate [34]. Due to the heterogeneous nature of concrete, the relationship between the DIF and strain rate has also been investigated with respect to various properties related to the material constituents, including mortar strength [35], aggregate type (*e.g.*, coral [36] or recycled [37] aggregate), aggregate morphology (*e.g.*, size [38] and shape [39]). Besides the interest in the macroscopic dynamic performance of concrete, advanced precision instruments, *e.g.*, X-ray computed tomography (XCT) [40] and ultra-high-speed imaging camera [41], have been widely used to investigate the microscopic effects. Nevertheless, analyses based on the available experimental data do not intuitively reflect the damage evolution of the internal material constituents. Direct evidence of the local response of concrete is needed to explain the macroscopic performance.

In SHPB tests, the specimen strain rate is often governed by the magnitude of the stress wave [30], and other factors associated with the impact wave, such as wave shape due to the presence of the pulse shaper between the striker and incident bars, have rarely been assessed in experiments. The main function of the pulse shaper in SHPB tests is to produce a longer and smoother ramp-like stress wave, characterised by the loading ramp rate, magnitude and duration of the stress wave [42]. As it is difficult to control the wave shape using a pulse shaper in experiments, this issue could be addressed using numerical approaches. It is shown that the stress wave impacting the incident bar can be prescribed in a waveform, allowing the wave shape to be defined in the SHPB model [43]. When different wave shapes are used to assess the dynamic response of concrete, they may also result in different ramp rates, the effects of which have been neglected and remain to be quantified.

To understand concrete responses under realistic dynamic stress conditions, various confined conditions have been taken into consideration. Confining pressure can be applied using active confinement methods, *e.g.*, liquid- or air-pressurised hydraulic vessels [44, 45], or passive confinement methods, *e.g.*, shrink-fit metal sleeve [46, 47]. Generally, the confinements can improve the concrete strength but cause lower strain rate sensitivity compared to unconfined cases [44-47]. Coupling dynamic loading with confinement is experimentally challenging, requiring a complex test setup and implementation. Consequently, it is difficult to maintain a constant confining pressure throughout the dynamic deformation of the concrete specimen, leading to an inaccurate estimation of the intended confining pressure. Numerical work on confined loading scenarios is quite limited and has mainly focused on developing well-calibrated models by comparing a single dynamic strength under unconfined and confined loading cases [48, 49] or under different confinement methods [21]. Systematic assessments of the overall and local responses of concrete under progressively increasing confinement levels are still lacking. For the abovementioned the loading ramp rate and confinement, their influences require further attention in SHPB-related studies.

Considering the different material phases of concrete, representing its internal structure using mesoscale models has been given particular attention. Generally, these models can be generated by either image-dependent approach, *i.e.*, XCT [50], or image-independent approaches, *i.e.*, computer based packing algorithms [51-53]. The former can be used to obtain the digital images of the concrete specimen and then convert them into the mesoscale structure of concrete [50]. Such an approach is highly affected by image resolution [54]. The image segmentation may fail to delineate some aggregate boundaries when encountering aggregate overlapping [54]. Alternatively, the latter makes it possible to overcome this issue by generating the spatial distribution of aggregates using various packing algorithms, such as take-in-place methods [55], random extension methods [56] and graphical methods [57]. These approaches can achieve a predefined packing density while simultaneously satisfying the specified Fuller curves for the aggregate size distribution. However, aggregates are represented using simplified and regular shapes, such as spheres, ellipsoids, or polyhedron [53], which fail to capture the realistic morphology of aggregates and may induce unrealistic stress concentrations compared with realistic shapes. Detecting aggregate-to-aggregate contact in packing is a challenging task for realistically shaped aggregates. For an effective approach, spherical harmonics [52] or Fourier descriptor-based methods [51, 58], combined with Voronoi tessellation for predefined spaces, have been adopted to generate realistic aggregate shapes within the cells, characterised by multiscale geometrical indices such as fractal dimension and relative roughness [51, 52].

Appropriate numerical methods and constitute laws need to be carefully considered for mesoscale models to reflect realistic dynamic fracture behaviour in concrete. Generally, numerical methods fall into two categories including discrete approaches and continuum approaches. Compared with the latter, the former requires extensive trial-and-error during calibration to determine the parameter values for the bonds between individual components, e.g., spheres in the discrete element method (DEM), as reported in some studies [59]. This issue can be avoided by using continuum modelling, which treats the system as a continuous medium rather than a collection of individual components [60]. The most representative continuum approach is the finite element method (FEM). Among the constitutive models implemented in FEM, such as concrete damaged plasticity (CDP) [53], cohesive interface element (CIE) methods [52], and the K&C model [61], some have been used to account for the dynamic nature of concrete and to predict its stress-strain behaviour under different strain rates. For the characterisation of material heterogeneity in mesoscale modelling, distinct material parameter values can be assigned to each material constituent, by assuming all have the same dynamic constitutive framework [62]. Alternatively, different constitutive models can be used for different material phases, such as combining the cohesive element method for the ITZ and CDP model for the mortar and aggregate [51, 63].

The well-established mesoscale modelling framework, considering morphological and material heterogeneities, has been employed to assess the dynamic response of concrete. These numerical works are generally conducted using two loading schemes, velocity and stress wave controlled. The first approach is to apply the velocity boundary conditions directly to the specimen [64]. The other approach is to use the impact wave in SHPB tests, generated by the striker impacting the incident bar [21], which has become more popular as it allows the investigation of the dynamic response at even higher strain rates. Relevant mesoscale modelling has been conducted to investigate the various interior and exterior influencing factors, such as low temperature [65], confined condition [21], specimen size [64] and shape, lightweight aggregate [66], aggregate solid fraction [67] and size [68]. Although damage or failure in different material phases can be visualised, such microscopic investigations are insufficient to explain macroscopic findings due to the lack of statistical and correlative analyses. As shown in the work of Liu et al. [51], which focuses on the quasi-static triaxial failure of concrete, statistical parameters obtained from PDF analyses of microscopic information can provide more evident support for the results. These model-based analyses must be properly calibrated and benchmarked [21], including the material parameters and internal contact model between post-failure segments. When velocity-controlled loading or impact waves are applied to the concrete model through a loading plate [64] or SHPB bars [21], the interaction with the concrete specimen is normally treated as frictional contact [21, 64]. Internal friction between and within different material phases, as an inherent property of concrete that is inaccessible in experiments, should also be incorporated into the model [51, 52]. However, its effects are often neglected in many studies and need to be quantified to enhance the understanding of the realistic responses of concrete material constituents to dynamic loading.

Despite based on the mesoscale structure of concrete, most numerical SHPB models emphasis on macroscopic dynamic responses, without providing sufficient information from microscopic analyses. During the loading process, the internal strain rate within the specimens exhibits temporospatial distributions due to the presence of material heterogeneities and interfaces [69]. Further, as additional microscopic information, local damage can effectively reflect the responses within concrete and explain the variation in its strength [63]. Both strain rate and damage aspects are rarely evaluated statistically or quantitively to provide direct evidence on the dynamic behaviour of concrete. For exploring the effects of loading ramp rate, internal friction, and confining pressure, a comprehensive numerical investigation of their respective effects is required, with particular attention given to analyses of internal strain rate and local damage to provide microscopic evidence supporting variations in the macroscopic dynamic response of concrete materials.

To address these research gaps discussed earlier, in this study, FEM modelling of SHPB tests is conducted to investigate the dynamic response of concrete. Here, concrete is represented as a three-phase mesoscale structure with realistic aggregates embedded in a homogeneous mortar matrix. The mesoscale model is used to explore the roles of internal strain rate and local damage in the DIF-strain rate relationship from different aspects, including loading ramp rate, internal friction, and confining pressure. The remainder of the paper is organised as follows. Section 2 presents the generation of a mesoscale concrete model with realistic aggregate shapes, followed by the implementations of strain-rate-dependent constitutive models and SHPB loading conditions. Model validations against experimental data are then conducted. Section 3 will assess each aspect by investigating the relationship between DIF and strain rate, through microscopic analyses of internal strain rates and damage states. Finally, conclusions are drawn in Section 4.

## 2. Methods

In this section, the mesoscale modelling of SHPB tests on concrete is conducted. The cylindrical concrete sample is represented as a mesoscale structure packed with realistically shaped aggregates Then, the material constitutive laws are described, combining the concrete damage plasticity model for the mortar and aggregates and the cohesive element method for the ITZ, capturing the dynamic properties of concrete. The loading conditions for modelling the SHPB test are presented. Based on the data processing, the validation of our numerical results against existing experimental data is provided.

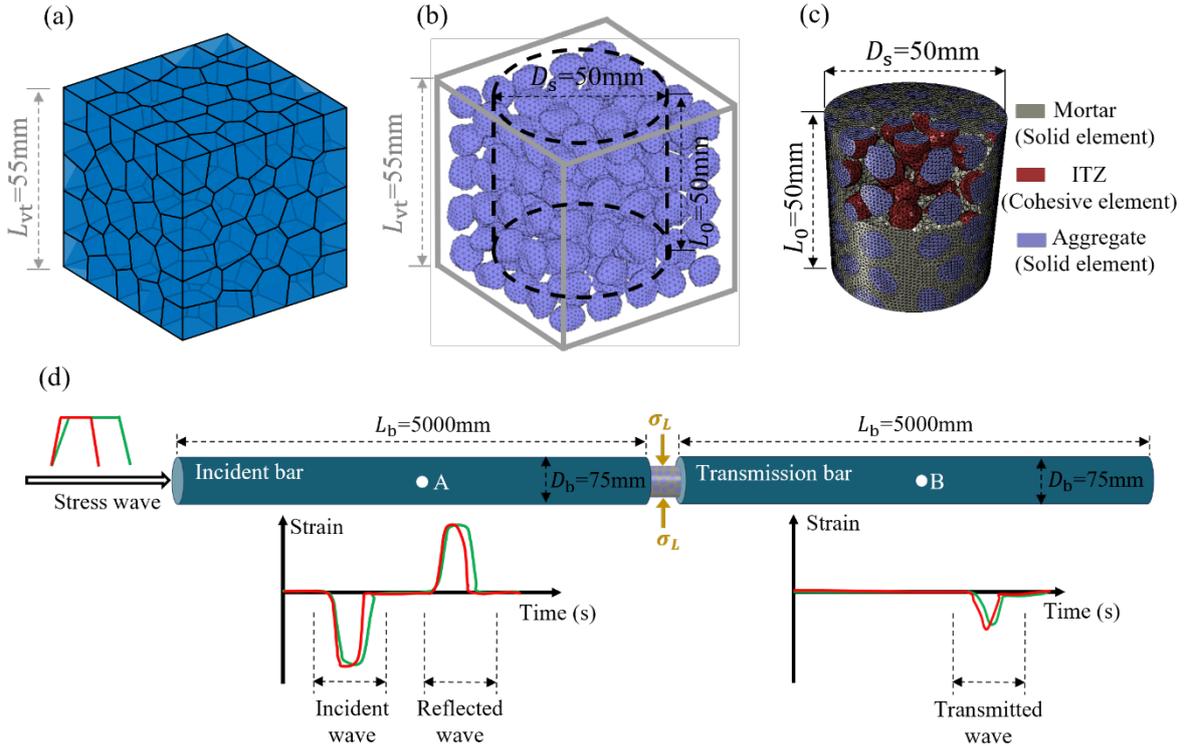

Figure 1. **Mesoscale modelling of SHPB tests on concrete**: (a) Typical Voronoi tessellation for packing processes; (b) Spatial distribution of irregularly shaped aggregates corresponding to the initial tessellation in (a); (c) A cylindrical specimen extracting from (b) containing three concrete phases and finite element meshes; and (d) Schematics of FE model of the SHPB test with a confining pressure of $\sigma_L$ and the wave records at points A and B.

### 2.1. Mesoscale sample generation

To generate the concrete model containing realistic aggregate shapes, this study uses the packing algorithms [51], *i.e.*, combination of Voronoi tessellation and the Fourier-descriptor-based methods. This approach has been improved by introducing the fractal dimension $F_d$ as a multiscale geometrical index to characterise the realistic aggregate shape. Given a target packing density, an appropriate size scaling of aggregates from their mass centres is also required to achieve a packing of realistically shaped aggregates without any contact or overlap within a predefined cubic space. As our early works have explored the effects of aggregate morphology, and the samples with $F_d = 2.3$ have been found to be the most representative case and only one geometrical model is used in all simulations due to the neglectable sample variation found previously.

The mesoscale structure of a cylindrical concrete specimen with a diameter of $D_s = 50$ mm and a length of $L_0 = 50$ mm is used for our numerical investigation. We first achieve aggregate packing within a predefined cubic space with a side length of 55 mm ($> L_0$), shown in Fig. 1(a), which is more than five times larger than the maximum aggregate size, as suggested by Naderi et al. [70]. The aggregates are then cut for generating the cylindrical concrete specimen. In the initial cubic space, 120 nearly mono-sized aggregates are generated using the adopted packing algorithm to achieve a packing density (or aggregate solid fraction, $SF_a$) of 30%. Generating realistic aggregate shapes with $F_d = 2.3$ in each cell produces the approximately mono-sized aggregate packing shown in Fig. 1(b), with a mean equivalent volume diameter of around 9.3 mm.

The aggregate surface is defined by 162 vertices and discretised into 320 nearly uniform triangular surficial meshes, which are directly generated using the published code from the work of Mollon and Zhao [58] that incorporates the concept of geodesic structures. The aggregate surfaces are then imported as solid geometries into the finite element software ABAQUS/Explicit. The solid aggregates are then cut using a cylindrical space with $D_s = 50$ mm and $L_0 = 50$ mm, whose centroid is aligned with the centre of the cubic space. As a result, 81 aggregates intersected by the cylindrical space are cut, while 39 aggregates remain entirely inside the domain. The mortar phase is then generated by filling the cylindrical space around these aggregates.

For meshing the cylindrical sample geometry directly cut from the initial cubical space, appropriate pre-treatment must first be applied to resolve the complex intersections among the elements, interfaces and exterior boundary. Cutting the aggregates with the cylindrical space may subdivide some triangular surface elements into small fragments, which induce excessively fine meshes around aggregate interfaces during the meshing process. This can be resolved by generating a more uniform mesh layout to replace the undesired fine meshes and prevent potential numerical errors. While preserving the geometry of the aggregate cuts, some vertices of the small mesh fragments need to be removed through trial and error. With this pre-treatment, the entire solid geometry of mortar and aggregate can be successfully discretised using tetrahedral elements with an edge size of approximately 1 mm, as shown in Fig. 1(c). More effective approaches are recommended for future work to avoid such meshing issues arising from aggregate cutting. The zero-thickness cohesive interface element (CIE) is used to represent the ITZ, following our previous work [51]. The meshed mesoscale structure of the cylindrical concrete specimen is shown in Fig. 1(c), containing around 418,000 solid elements for the mortar and aggregates and around 34,000 CIEs for the ITZs. The size of tetrahedral element around 1 mm, approximately 1/10 of the characteristic aggregate size, has been commonly adopted in other studies on concrete fracture and proven to be sufficient for ensuring simulation accuracy [51, 70, 71].

## 2.2 Constitutive models for dynamic fracture

Different constitutive laws for concrete materials are considered to simulate dynamic fracture. Here, the concrete damage plasticity (CDP) model is used for both mortar and aggregate, with respective sets of material parameters. For the ITZs, the traction-separation law is used to describe the mechanical response of the CIEs. The key aspects of the constitutive laws and the relevant strain rate effects are also discussed in this section.

Table 1: Material properties of concrete constituents.

| Constituents | Material property | Value |
| --- | --- | --- |
| Mortar/Aggregate (Concrete damage plasticity, CDP) | Density (kg/m$^3$) | 2200/2600 |
| | Young's modulus (GPa) | 30/70 |
| | Poisson's ratio | 0.2/0.2 |
| | Ultimate strength in compression (MPa) | 45/80 |
| | Ultimate strength in tension (MPa) | 4/10 |
| | Fracture energy, $GFI_{ts}$ (N/m) | 50/60 |
| ITZ (Cohesive interface element, CIE) | Initial stiffness (N/m$^3$) | $1 \times 10^{13}$ |
| | Normal strength (MPa) | 2.4 |
| | Shear strength (MPa) | 7.2 |
| | Fracture energy (N/m) | 30 |
| Steel SHPB bar | Density (kg/m$^3$) | 7850 |
| | Young's modulus (GPa) | 200 |
| | Poisson's ratio | 0.3 |

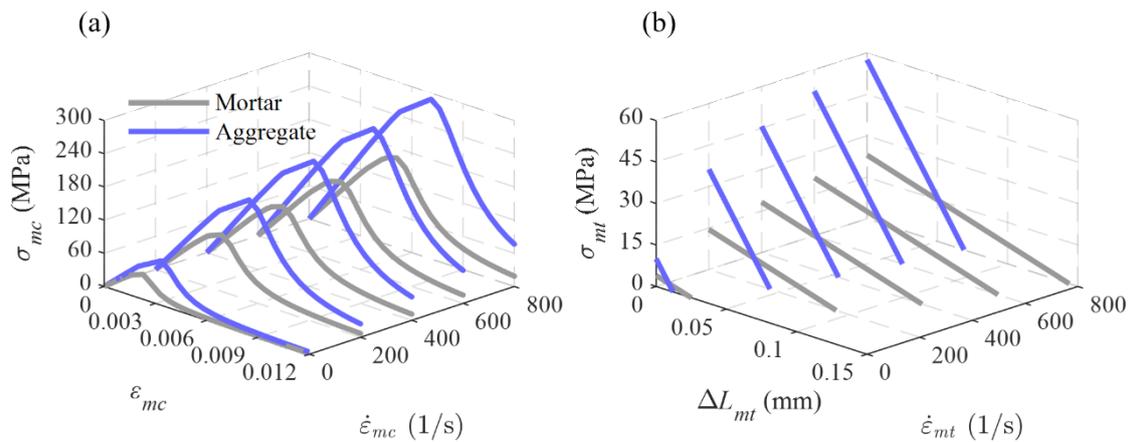

Figure 2. **Dynamic responses of the material phases**: (a) Compressive stress ($\sigma_{mc}$) -strain ($\varepsilon_{mc}$) under varying compressive strain rates $\dot{\varepsilon}_{mc}$, and (b) Post-peak tensile stress ($\sigma_{mt}$)–crack opening ($\Delta L_{mt}$) under varying tensile strain rates, $\dot{\varepsilon}_{mt}$.

**2.2.1 Concrete damage plasticity (CDP) model**

The CDP model defines compression and tension as the two primary failure mechanisms, based on the concepts of isotropic damage together with isotropic plasticity to describe the inelastic behaviour of concrete [72]. The compressive stress-strain relationship is defined using the FIB Model Code 2010 (FIB-MC2010) [32], in which the stress ($\sigma_{mc}$)-strain ($\varepsilon_{mc}$) curve is linear elastic up to the onset of damage, followed by strain-hardening and strain-softening stages [32, 64]. For tension, specifying the stress-strain curve may lead to mesh-size dependency in the numerical results [73, 74]. To address this issue, the tensile behaviour can alternatively be defined by directly specifying the tensile strength and fracture energy in the CDP model to characterise the tensile stress ($\sigma_{mt}$) - crack opening ($\Delta L_{mt}$) response [73, 74]. The corresponding material parameters have been previously calibrated with sets of experimental data as summarised in Table 1 [51]. In addition, five other parameters of the CDP model, which are dilation angle, plastic potential eccentricity, ratio of biaxial to uniaxial compressive strength, ratio of the second stress invariant on the tensile meridian to that on the compressive meridian, and viscosity parameter, are taken to be the same for both the mortar and aggregate phases, with values of $35°$, 0.1, 1.16, 0.667, and 0.0005, respectively [51].

**2.2.2 Cohesive element method for ITZ**

A bilinear traction-separation law is employed to model the response of the cohesive elements, exhibiting linear behaviour after damage initiation, followed by energy-based linear softening [75]. Damage initiates when the stresses in the normal and shear directions reach their critical values, as defined by a quadratic stress criterion. Once damage is initiated, the tractions gradually decrease with increasing separation, representing softening and energy dissipation through fracture. The fracture energy corresponds to the total energy required to fully separate the CIE. When the separation reaches the ultimate fracture displacement, the CIE is completely damaged. Benzeggagh-Kenane (BK) criterion [76] is adopted to account for mixed-mode behaviour. The material parameters for the CIE [51] have been previously calibrated as in Table 1.

**2.2.3 Strain rate effects**

The constitutive models for concrete materials should be adapted to respond to the real-time strain rate at each incremental step to accurately account for the dynamic response of concrete. This rate-dependent response is reflected in mechanical properties such as compressive and tensile strength, elastic modulus, and strain at peak stress [62, 73], several of which are well defined in FIB-MC2010 [32]. Among these properties, the influence of strain rate on concrete strength is the most significant and is therefore taken as the main consideration in mesoscale modelling [62, 73]. This can be achieved by adapting strengths of material constituents to the real-time strain rates, assuming that they exhibit a similar strain rate behaviour as the overall concrete strength [62, 73]. The compressive and tensile strengths of both mortar and aggregate are defined as strain-rate dependent based on the formulas in FIB-MC2010 [32], as shown below:

Compression:

$$\frac{\sigma_{mc,d}}{\sigma_{mc,s}} = \left(\frac{\dot{\varepsilon}_{mc}}{\dot{\varepsilon}_{mc,s}}\right)^{0.014}, \dot{\varepsilon}_{mc} \leq 30 \text{ s}^{-1} \tag{1}$$

$$\frac{\sigma_{mc,d}}{\sigma_{mc,s}} = 0.012\left(\frac{\dot{\varepsilon}_{mc}}{\dot{\varepsilon}_{mc,s}}\right)^{1/3}, \dot{\varepsilon}_{mc} > 30 \text{ s}^{-1} \tag{2}$$

Tension:

$$\frac{\sigma_{mt,d}}{\sigma_{mt,s}} = \left(\frac{\dot{\varepsilon}_{mt}}{\dot{\varepsilon}_{mt,s}}\right)^{0.018}, \dot{\varepsilon}_{mt} \leq 10 \text{ s}^{-1} \tag{3}$$

$$\frac{\sigma_{mt,d}}{\sigma_{mt,s}} = 0.0062\left(\frac{\dot{\varepsilon}_{mt}}{\dot{\varepsilon}_{mt,s}}\right)^{1/3}, \dot{\varepsilon}_{mt} > 10 \text{ s}^{-1}, \tag{4}$$

where $\sigma_{mc,s}$ and $\sigma_{mt,s}$ are the static compressive and tensile strengths (MPa), $\dot{\varepsilon}_{mc,s} = 30 \times 10^{-6} \text{s}^{-1}$ and $\dot{\varepsilon}_{mt,s} = 1 \times 10^{-6} \text{s}^{-1}$ are the reference static strain rates in compression and tension, $\sigma_{mc,d}$ and $\sigma_{mt,d}$ are dynamic compressive and tensile strength (MPa) at their respective strain rate $\dot{\varepsilon}_{mc}$ and $\dot{\varepsilon}_{mt}$ (s$^{-1}$). The subscription, $m$, indicates the corresponding material constituent.

As suggested by Snozzi et al. [77], the strain-rate effects on the tensile fracture energy are defined in this study to ensure that the tensile strength and crack opening increase proportionally with increasing strain rate. For simplicity, the fracture energy $\text{GFI}_{mt,d}$ at different strain rates can be calculated by replacing $\frac{\sigma_{mt,d}}{\sigma_{mt,s}}$ with $\frac{\text{GFI}_{mt,d}}{\text{GFI}_{mt,s}}$ in Eqs.(3) and (4), where $\text{GFI}_{mt,s}$ is fracture energy (N/m) at the reference static strain rate.

For the remaining material parameters in Table 1, the strain rate effects are considered negligible in the CDP model. Eventually, the compressive stress–strain curves, tensile strengths, and fracture energy for mortar and aggregate can be adapted to increasing strain rates and provided as input to the implemented CDP model. Examples under several compressive and tensile strain rates are shown in Fig. 2, where the post-peak tensile stress ($\sigma_{mt}$)–crack opening ($\Delta L_{mt}$) in Fig. 2(b) is used to demonstrate the effects of strain rate on tensile strength and fracture energy.

Finally, the ITZ is modelled using zero-thickness CIEs and does not possess measurable strain fields, unlike solid phases of mortar or aggregate. The concept of strain rate, defined based on deformation over a finite length and time, is not physically meaningful for zero-thickness CIEs. Although some efforts have been dedicated to developing traction–separation laws that incorporate opening or sliding rate effects for CIEs [78, 79], such a rate dependency can be neglected for our model, due to insufficient experimental observations reported about dynamic response of ITZ [62, 80, 81].

## 2.3 Loading conditions and wave signal processing

### 2.3.1 Loading conditions

The modelling of the SHPB test on concrete specimens is conducted to investigate the dynamic behaviour of concrete, as shown in Fig.1(d). The incident and transmission bars have a diameter of $D_b = 80$ mm and a length of $L_b = 5000$ mm, and both are meshed with around 4 mm sized hexahedral solid elements. The bars are made of steel, with material properties shown in Table 1. The cylindrical concrete specimen is positioned between the two bars, and the contact between the bar and specimen surfaces is defined as hard contact with a friction coefficient of 0.5. In confined cases, the confining pressure $\sigma_L$ is kept below half of the uniaxial compressive strength of concrete under quasi-static conditions. Due to its efficiency in handling contacts, ABAQUS/Explicit is employed as the dynamic solver for all modelling cases.

The loading scheme is conducted in two sequential steps without any additional step coupling. In the first stage, the confining pressure is applied to the lateral surface of the cylindrical specimen and linearly increased to the specified value within 0.001 s. It then remains constant until the end of the entire simulation. In the second stage (with a duration of 0.0023 s), the stress wave impacting the incident bar is used by a prescribed loading amplitude that defines the variation of stress with time. A trapezoidal loading amplitude is employed to reduce excessive oscillations associated with instantaneous loading, in comparison with other shapes, as noted by Ayhan et al. [43]. Details of the loading amplitude, such as the maximum stress and duration, are specified in the following sections for different purposes of interest. The loading ramp rate, describing the increase rate of the stress at the initial rising stage of trapezoidal loading, has been controlled for highlighting the sensitivity of the stress wave shape. To determine the DIF, quasi-static fracture of concrete also needs to be modelled, as provided in Appendix A.

### 2.3.2 Wave signal processing

As shown in Fig. 1(d), points A and B are the midpoints of the incident and transmitted bars, respectively, where the strain waves $\varepsilon_w(t)$ are recorded over time $t$ to calculate the dynamic stress–strain response of the concrete specimen. When the compressive stress wave impacts the incident bar, it travels along the bar and is recorded at point A as the incident strain wave, $\varepsilon_{w,i}(t)$. Upon reaching the interface between the incident bar and the cylindrical specimen, a portion of the incident wave is reflected back into the incident bar and recorded at point A as the reflected strain wave, $\varepsilon_{w,r}(t)$. The remaining portion is transmitted through the specimen and travels back and forth between its two end surfaces of the specimen over several cycles. Finally, the stress wave exits the specimen, propagates along the transmitted bar, and is recorded at point B as the transmitted strain wave, $\varepsilon_{w,t}(t)$. As an example, the recorded $\varepsilon_{w,i}(t)$, $\varepsilon_{w,r}(t)$, and $\varepsilon_{w,t}(t)$ at $\sigma_{ld,max} = 50$ MPa are presented in Fig. 3(g).

According to one-dimensional wave theory [82], the time histories of strain, $\varepsilon(t)$, stress, $\sigma(t)$, and strain rate, $\dot{\varepsilon}(t)$, of the specimen can be calculated as follows:

$$\varepsilon(t) = \frac{C_b}{L_0} \int_0^t [\varepsilon_{w,i}(t) - \varepsilon_{w,r}(t) - \varepsilon_{w,t}(t)]\, dt, \tag{5}$$

$$\sigma(t) = \frac{E_b A_b [\varepsilon_{w,i}(t) + \varepsilon_{w,r}(t) + \varepsilon_{w,t}(t)]}{2 A_s}, \tag{6}$$

$$\dot{\varepsilon}(t) = \frac{C_b}{L_0} [\varepsilon_{w,i}(t) - \varepsilon_{w,r}(t) - \varepsilon_{w,t}(t)], \tag{7}$$

where $C_b$ is the wave velocity (mm/s) in the bar and can be calculated as $\sqrt{E_b/p_b}$, $E_b$ and $p_b$ are the elastic modulus (N/mm$^2$) and density (kg/mm$^3$) of bar, respectively, $L_0$ is the initial length (mm) of the cylindrical concrete specimen, and $A_b$ and $A_s$ are the cross section area (mm$^2$) of the bar and cylindrical concrete specimen, respectively. The strain rate of the concrete specimen is generally not constant throughout the dynamic loading period. For a representative value, the strain rate at the peak stress, $\dot{\varepsilon}_p$, is adopted in this study, as particular emphasis is placed on the dynamic strength in following sections. For the quasi-static condition, the stress is defined as $\sigma = F/A$, where $F$ is the sum of the nodal reaction forces (N) on the top plate, and $A$ is the initial cross-sectional area (mm$^2$) of the cylindrical specimen. The strain is calculated as the ratio of the nodal displacement (mm) to the initial length (mm) of the specimen. Although the velocity-controlled loading method for the quasi-static case involves a short acceleration phase from zero to the prescribed velocity, this brief period can be neglected, allowing the loading process to be considered as occurring at a constant velocity. Thus, the strain rate at the peak stress for the quasi-static case is 1/s, obtained by dividing the applied maximum velocity (50 mm/s) by the initial specimen length ($L_0 = 50$ mm), in accordance with the approach by Liu et al. [62].

## 2.4 Validation

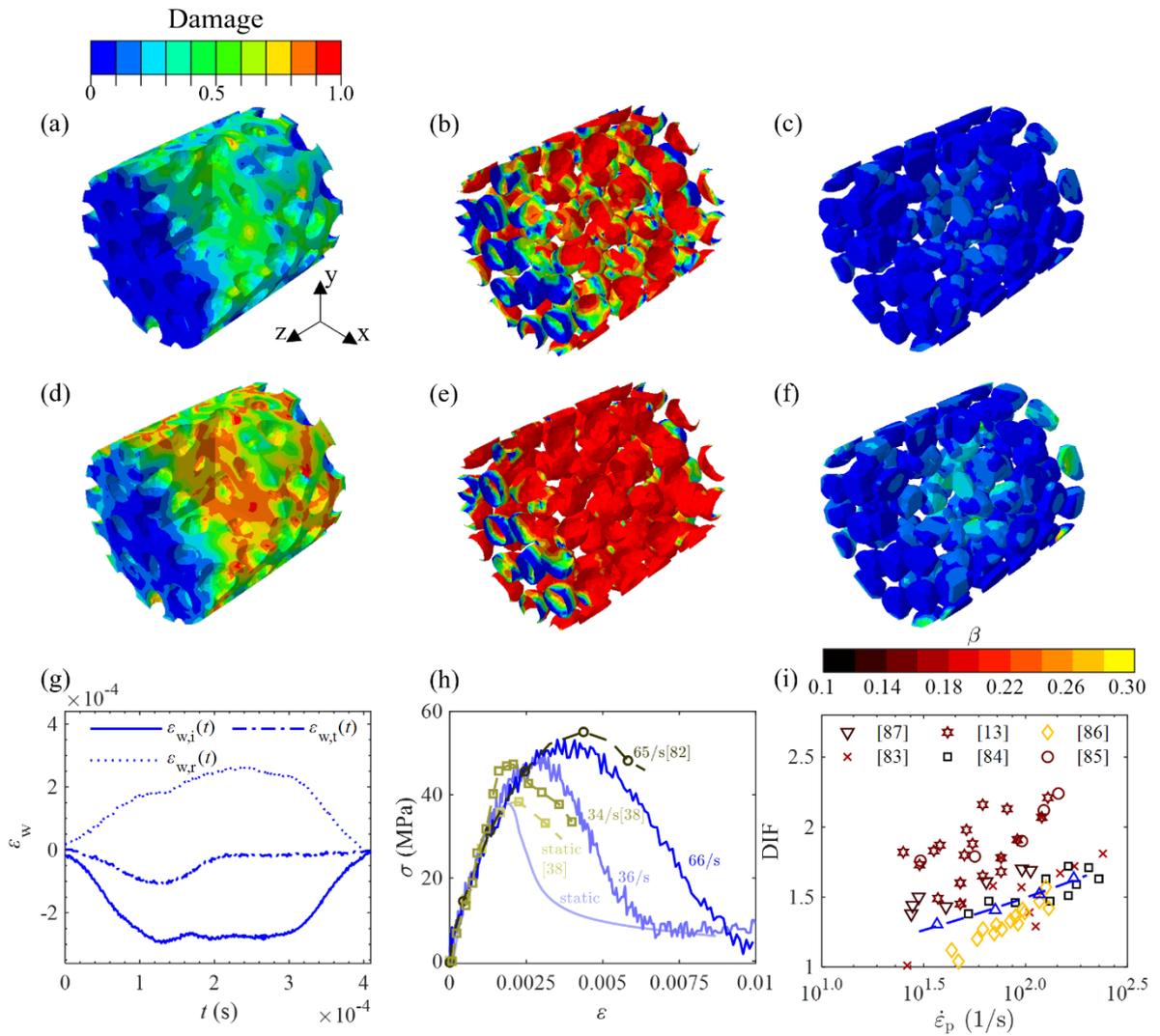

Figure 3. **Model validation:** (a-f) Damage field distributions of mortar, ITZ, and aggregate at the peak dynamic strength, where (a-c) and (d-f) correspond to maximum stress of trapezoidal loading amplitude, $\sigma_{\text{ld,max}}$ = 50 and 250 MPa, respectively; (g) A typical incident $\varepsilon_{\text{w,i}}(t)$, reflected $\varepsilon_{\text{w,r}}(t)$, and transmitted $\varepsilon_{\text{w,t}}(t)$ waves with $\sigma_{\text{ld,max}}$ = 50 MPa; (h) Stress ($\sigma$)-strain ($\varepsilon$) responses of the concrete specimen under the quasi-static and dynamic cases, with comparable strain rates labelled using different gradual colours for numerical and experimental data; and (i) The dynamic increase factor (DIF) vs the strain rate of the specimen ($\dot{\varepsilon}_{\text{p}}$), where the dashed line represents the fitting to the numerical results, and the colour bar above indicates the corresponding $\beta$ values of the experimental datasets, represented by marker shape in different colours.

To validate our results, this section focuses only on the rate effects of the unconfined cases. Internal friction between material phases within the concrete model is one of the key aspects examined in Section 3.2. Here, the internal friction coefficient $f_{\text{e}} = 0.1$ is considered. The trapezoidal loading amplitude for the impact wave is prescribed such that the stress increases linearly to its maximum within 0.0001 s, remains constant for 0.0002 s, and then decreases linearly to zero within 0.0001 s. The maximum stress for the trapezoidal loading amplitude, $\sigma_{\text{ld,max}}$, is specified as 50, 90, 150, and 250 MPa to evaluate the dynamic response of concrete under increasing strain rates. These settings in the SHPB model are adopted for following investigations unless otherwise specified for a particular purpose.

In Fig. 3(h), the stress-strain curves are obtained for the quasi-static case and for the dynamic cases with $\sigma_{\text{ld,max}}$ of 50 and 90 MPa, which yield $\dot{\varepsilon}_p$ values of 1, 36, and 66/s, respectively. At higher strain rates, the stress-strain behaviour becomes more ductile, with a significant improvement in concrete strength. To achieve effective validation, the experimental data used for comparison must be obtained at comparable strain rates and from specimen the same slenderness ratio of specimen ($\frac{L_0}{D_s} = 1$). Our results capture the general trend of the limited available stress–strain behaviour reported in experimental studies at different strain rates. For the dynamic cases with higher $\sigma_{\text{ld,max}}$ values of 150 and 250 MPa, which yield higher strain rates of $\dot{\varepsilon}_p = 116$ and 172 /s, we acknowledge the lack of relevant experimental data as a limitation for further validating the model through stress-strain comparisons, in particular at the high strain rate regime. Alternatively, the dynamic increase in concrete strength, as a key concern in dynamic response of concrete, will be compared with existing studies in the following.

The dynamic increase factor (DIF), defined as the ratio of dynamic to static compressive strength of concrete, is computed under varying strain rates, as shown in Fig. 3(i). As the DIF increases with increasing strain rate, the damage in the mortar, ITZ and aggregates becomes more severe, as indicated by comparing Fig. 3(a-c) with (d-f), because higher stresses are imposed on the elements, accelerating the respective damage. Our numerical results are plotted against experimental data that consider various influencing factors, such as normal concrete [83-85], cement and aggregate types [86], additives [87], and moisture content [13]. The fitting to the numerical results is proposed in a form similar to Eq. (2) and achieves an excellent coefficient of determination $R^2 = 0.99$:

$$\text{DIF} = \frac{\sigma_{d,p}}{\sigma_{s,p}} = \alpha \left(\frac{\dot{\varepsilon}_p}{\dot{\varepsilon}_{p,0}}\right)^\beta, \tag{8}$$

$$\frac{\text{DIF}}{\text{DIF}_1} = \frac{\alpha \left(\frac{\dot{\varepsilon}_p}{\dot{\varepsilon}_{p,0}}\right)^\beta}{\alpha \left(\frac{\dot{\varepsilon}_{p,1}}{\dot{\varepsilon}_{p,0}}\right)^\beta} = \left(\frac{\dot{\varepsilon}_p}{\dot{\varepsilon}_{p,1}}\right)^\beta, \tag{9}$$

where $\sigma_{d,p}$ and $\sigma_{s,p}$ are the dynamic and static strengths of concrete, respectively; $\dot{\varepsilon}_p$ is the strain rate; $\dot{\varepsilon}_{p,0}$ is reference quasi-static strain rate, 1/s, $\alpha = 0.7613$ is the fitting parameter that defines the baseline of the DIF, and $\beta = 0.1466$ is the fitting parameter that characterises the slope of the DIF-$\dot{\varepsilon}_p$. Due to the missing value of $\dot{\varepsilon}_{p,0}$ in some experimental datasets, Eq. (8) cannot be applied directly to obtain their $\beta$ values, which are also of particular importance to our validation. For this concern, we normalise the DIF by $\text{DIF}_1$, the first DIF value in the dataset, thereby eliminating the need for $\dot{\varepsilon}_{p,0}$. This enables a fitting with $R^2 > 0.9$ to the experimental data to obtain the $\beta$ values using Eq. (9), where $\dot{\varepsilon}_{p,1}$ is the first strain rate of each experimental dataset. In Fig. 3(i), the numerically obtained DIF values fall within the experimental range, with an acceptable slope when comparing the $\beta$ values with those of the experimental results shown in the colour bar. This successful validation confirms the reliability of our SHPB models and enables the following investigations and analyses.

# 3. Results and Discussion

Under dynamic loading, both macroscopic and microscopic assessments of concrete are conducted, with particular focus on the effects of loading ramp rate, internal friction, and confining pressure. All these analyses emphases on exploring the role of internal strain rate and damage on DIF through quantitative and statistical analyses of obtained results.

## 3.1. Loading ramp rate

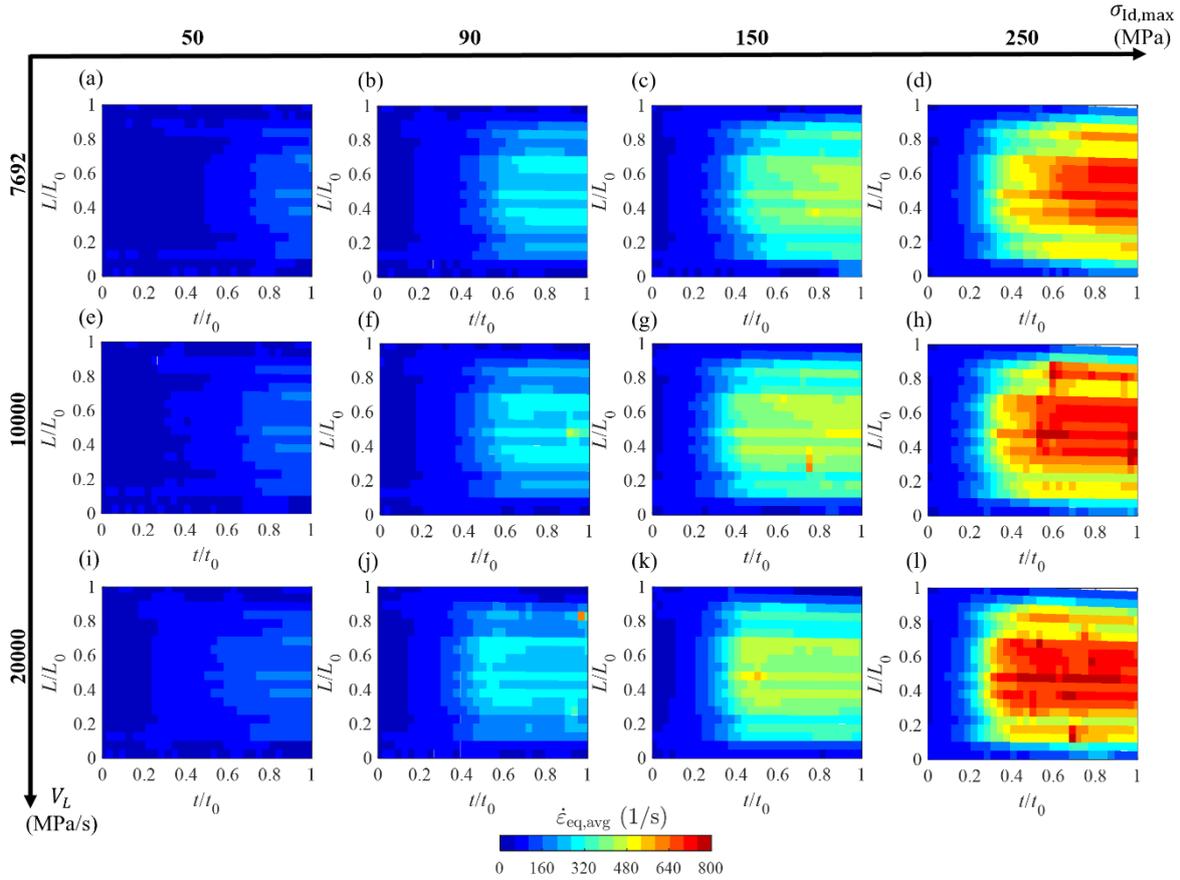

Figure 4: Time evolution of the profile of average internal strain rate, $\dot{\varepsilon}_{\text{eq,avg}}$, along the length deformed specimen, $L/L_0$, for different maximum stresses of the trapezoidal loading amplitudes, $\sigma_{\text{ld,max}}$, and loading ramp rate, $V_L$.

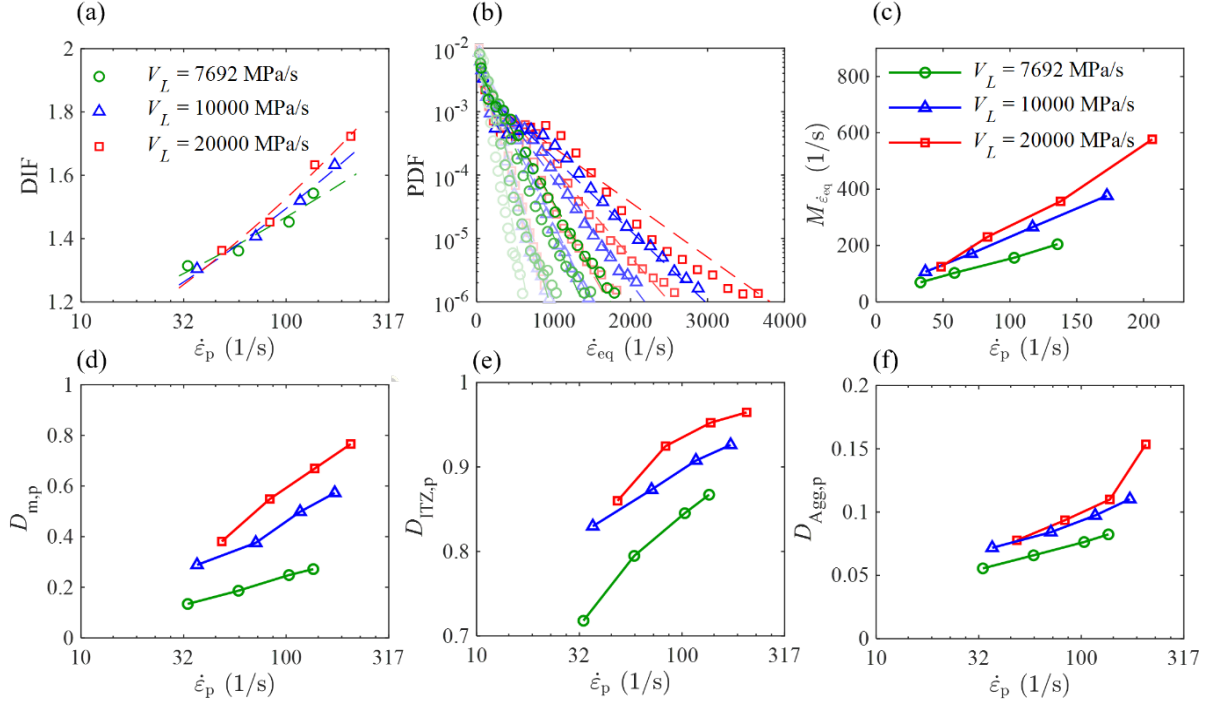

Figure 5. **Effects of loading ramp rate, $V_L$, on the macro- and micro- responses of concrete**: (a) DIF versus $\dot{\varepsilon}_p$ for different $V_L$, with dashed lines denoting the fits using the Eq.(8); (b) PDFs of the internal equivalent strain rate, $\dot{\varepsilon}_{eq}$, for different $\sigma_{ld,max}$ and $V_L$, using the same legend as in (a), with dashed lines denoting exponential fits and colours from light to dark indicating increasing $\dot{\varepsilon}_p$; (c) Mean value, $M_{\dot{\varepsilon}_{eq}}$, from exponential fitting of the PDFs; and (d-f) Damage of mortar, $D_{m,p}$, ITZ, $D_{ITZ,p}$, and aggregate, $D_{Agg,p}$, for the DIF, using the same legend in (c).

Here, we focus on the effect of the loading ramp rate, $V_L$, resulting from different pulse shapers adopted in different sets of SHPB tests. Our studied cases include maximum stress of $\sigma_{ld,max} = $ 50, 90, 150, and 250 MPa, and the stress magnitude has been kept for 0.0002 s. To vary the loading ramp rate, we include the cases of $V_L = $ 7,692, 10,000 and 20,000 MPa/s in the initial rising stage. After stress waves are recorded, the example with $\sigma_{ld,max} = $ 90 MPa in Fig. B.1 of Appendix B shows that a higher $V_L$ can result in a faster increase of the incident waves over the initial rising stage, thereby demonstrating the achievement of precise control over $V_L$. Using the Eqs. (5-6), stress ($\sigma$)-strain ($\varepsilon$) curves are calculated in Fig. B.2(b) of Appendix B, showing dependences on $\sigma_{ld,max}$ and $V_L$. Both effects on the concrete strength have been mainly focused by computing the DIF versus $\dot{\varepsilon}_p$ across various $V_L$ in Fig. 5(a). Using the Eq. (8), the $\beta$ values are measured as 0.113, 0.147, and 0.171 for $V_L = $ 7692, 10000 and 20000 MPa/s, respectively. A higher $V_L$ starts to lead to a higher DIF when $\dot{\varepsilon}_p$ increases to a sufficiently large value in the range of $\dot{\varepsilon}_p \in [100,200]$ /s, and eventually results in a more significant increase in DIF with increasing $\dot{\varepsilon}_p$, as reflected by the larger $\beta$ values.

We further explore the microscopic information, *i.e.*, internal strain rate and local damage, emphasising on the inherent heterogeneities and their implications on the macroscopic observations.

For the temporospatial variations of internal strain rate during dynamic loading, we approach this in a manner similar to that used by Blatny et al. [69], who quantified the evolution of the internal strain rate profile along the height of 2D porous media to successfully illustrate compaction patterns. The internal strain rate for each element, $\dot{\varepsilon}_{eq}$, available for mortar and aggregate elements, is calculated based on the second invariant of the strain rate tensor, as explicitly shown below:

$$\dot{\varepsilon}_{eq} = \sqrt{[\dot{\varepsilon}_{xx}^2 + \dot{\varepsilon}_{yy}^2 + \dot{\varepsilon}_{zz}^2 + \frac{1}{2}(\dot{\varepsilon}_{xy}^2 + \dot{\varepsilon}_{xz}^2 + \dot{\varepsilon}_{yz}^2)]}, \tag{10}$$

where $\dot{\varepsilon}_{xx}$, $\dot{\varepsilon}_{yy}$, $\dot{\varepsilon}_{zz}$, $\dot{\varepsilon}_{xy}$, $\dot{\varepsilon}_{xz}$, and $\dot{\varepsilon}_{yz}$ are the strain rate components. We focus only on the time duration of $t_0 = 0.0003$ s from the initial time $L_0/C_b$ of stress wave reaching the specimen. The $\dot{\varepsilon}_{eq}$ values are obtained for 50 time points $t$, evenly distributed over the duration $t_0$. At each normalised time point $t/t_0$, we define a moving window with a fixed thickness of 5 mm along the loading direction. As this window moves in an initial step of 2.5 mm mapped along the deformed specimen length, the average internal strain rate of all solid elements within the window, denoted as $\dot{\varepsilon}_{eq,avg}$, is calculated at varying $L/L_0$, where $L$ is the centroid position of the window and $L_0$ is the original specimen length. Eventually, the profile of $\dot{\varepsilon}_{eq,avg}$ along $L/L_0$ can be obtained for all 50 time points to form the heatmap, as shown in Fig. 4. In each case, as $t/t_0$ increases, the stress acting on the solid elements gradually increases, resulting in higher values of $\dot{\varepsilon}_{eq,avg}$ along $L/L_0$.

The evolution of the $\dot{\varepsilon}_{eq,avg}$ profile in Fig. 4 shows effects of the applied maximum stress in the trapezoidal loading amplitude, $\sigma_{ld,max}$, and loading ramp rate, $V_L$. With higher $\sigma_{ld,max}$, higher stresses imposed on the elements can promote deformation and induce the evolution of the higher $\dot{\varepsilon}_{eq,avg}$ profile. Comparison of Fig. 4(e-h) or (i-l) with (a-d) shows that a higher $V_L$ produces a more rapid stress increase, leading to a higher $\dot{\varepsilon}_{eq,avg}$ profile at an earlier time. Furthermore, this effect becomes more pronounced at higher $\sigma_{ld,max}$, which can further intensify the stress increase. These observations can be more clearly found in the central region than near the two ends, which experience stronger deformation constraints from the incident and transmission bars. These visualisations can present overviews of the internal strain rate throughout the entire deformation process, and further direct and quantitative evidences at the peak are required to explain the results of DIF-$\dot{\varepsilon}_p$.

The probability density function (PDF) is employed for the statistical analysis of $\dot{\varepsilon}_{eq}$ for all mortar and aggregate elements at the peak stress. The PDF is calculated based on the element volume to highlight the effects of $V_L$ and $\sigma_{ld,max}$ by the semi-log plots, as shown in Fig. 5(b). To explicitly describe these variations in the PDFs, the exponential fitting function, $f_{ex}(\dot{\varepsilon}_{eq})$, is applied to effectively capture the overall trend of the distributions, as shown below:

$$f_{\text{ex}}(\dot{\varepsilon}_{\text{eq}}) = \lambda e^{-\lambda \dot{\varepsilon}_{\text{eq}}}, \tag{11}$$

$$M_{\dot{\varepsilon}_{\text{eq}}} = \frac{1}{\lambda}, \tag{12}$$

where $\lambda$ is the fitting parameter of the exponential distribution. Using this fitting, Eq. (12) is applied to compute the mean value, $M_{\dot{\varepsilon}_{\text{eq}}}$ to characterise the central tendency of the PDFs.

The PDFs and $M_{\dot{\varepsilon}_{\text{eq}}}$ are shown in Fig. 5(b) and (c), respectively. Increasing $\sigma_{\text{ld,max}}$ and $V_L$ broadens the distribution toward higher $\dot{\varepsilon}_{\text{eq}}$ values, with more elements attaining larger $\dot{\varepsilon}_{\text{eq}}$. ultimately leading to higher $M_{\dot{\varepsilon}_{\text{eq}}}$. These tendencies in PDF and $M_{\dot{\varepsilon}_{\text{eq}}}$ shows similar effects of $\sigma_{\text{ld,max}}$ and $V_L$ to those observed in Fig. 4, confirming the existence of the similar underlying mechanism governing concrete strength. The enhancement of $M_{\dot{\varepsilon}_{\text{eq}}}$ due to increased $\sigma_{\text{ld,max}}$ or $V_L$ suggests that a greater fraction of elements undergo elevated local strain rates, which in turn promotes overall material strengthening and results in a higher DIF. Increasing $\sigma_{\text{ld,max}}$ intensifies the enhancing effect of higher $V_L$ on $M_{\dot{\varepsilon}_{\text{eq}}}$, leading to more material elements undergo high-strain-rate deformation. Consequently, a higher $V_L$ leads to a more pronounced increase in $M_{\dot{\varepsilon}_{\text{eq}}}$ with respect to $\dot{\varepsilon}_{\text{p}}$, reflecting the observed higher slope of the DIF-$\dot{\varepsilon}_{\text{p}}$ in Fig. 5(a).

Further, the damages of mortar, ITZ, and aggregate are expressed as $D_{\text{m,p}}$, $D_{\text{ITZ,p}}$, and $D_{\text{Agg,p}}$, respectively, and can be quantified as follows:

$$\frac{\sum_{i=1}^{N} d_i V_i}{\sum_{i=1}^{N} V_i} \text{ or } \frac{\sum_{i=1}^{N} d_i S_i}{\sum_{i=1}^{N} S_i}, \tag{13}$$

where $d_i$ is the damage of the $i$-th element, $V_i$ is the volume of the $i$-th solid element (for mortar or aggregate), and $S_i$ is the area of the $i$-th CIE (for the ITZ). The variations of $D_{\text{m,p}}$, $D_{\text{ITZ,p}}$, and $D_{\text{Agg,p}}$ with $\dot{\varepsilon}_{\text{p}}$ for each $V_L$ are computed in Fig. 5(d-f), respectively.

When the DIF is enhanced at higher $\dot{\varepsilon}_{\text{p}}$, greater damage develops within the different material phases, as the higher stresses acting on the elements due to the increase in $\sigma_{\text{ld,max}}$ can promote fractures. At higher $V_L$, larger instantaneous stresses are induced, promoting fracture in all material phases, as evidenced by the higher values of $D_{\text{m,p}}$, $D_{\text{ITZ,p}}$, and $D_{\text{Agg,p}}$, generally contributing to a higher DIF. With an increase in $\dot{\varepsilon}_{\text{p}}$, the effect of $V_L$ on damage exhibits inconsistent trends across different material phases. For the ITZ, this effect becomes less significant. This is mainly because, as the weakest zone, the ITZ is more susceptible to damage and gradually reaches a fully damaged state with increasing $\sigma_{\text{ld,max}}$, resulting in reducing differences in $D_{\text{ITZ,p}}$ among different $V_L$, although a higher $V_L$ still induces higher stress in the elements. In contrast, increasing $\sigma_{\text{ld,max}}$ amplifies the effect of $V_L$ on $D_{\text{m,p}}$ and $D_{\text{Agg,p}}$, results in more pronounced increases in $D_{\text{m,p}}$ and $D_{\text{Agg,p}}$ at higher $V_L$. This indicates that fracture in the mortar and aggregate, caused by higher $V_L$, can be further promoted by increasing $\sigma_{\text{ld,max}}$, ultimately leading to a greater enhancement of the DIF with increasing $\dot{\varepsilon}_{\text{p}}$.

In the following sections, a trapezoidal loading amplitude with $V_L = 10000$ MPa/s is retained to investigate the additional effects of internal friction and confining pressure. For each effect, the DIF-$\dot{\varepsilon}_p$ remains the main focuses in characterising the overall response of concrete. The same approach to analyse the internal strain rate and local damage will be applied to explore the variations in the DIF-$\dot{\varepsilon}_p$. Thus, the macro- and microscopic responses of concrete will be presented in a format consistent with that of Fig. 5. The following discussion focuses only information specific to how internal friction and confining pressure affect the DIF and the slope of DIF-$\dot{\varepsilon}_p$.

## 3.2. Internal friction

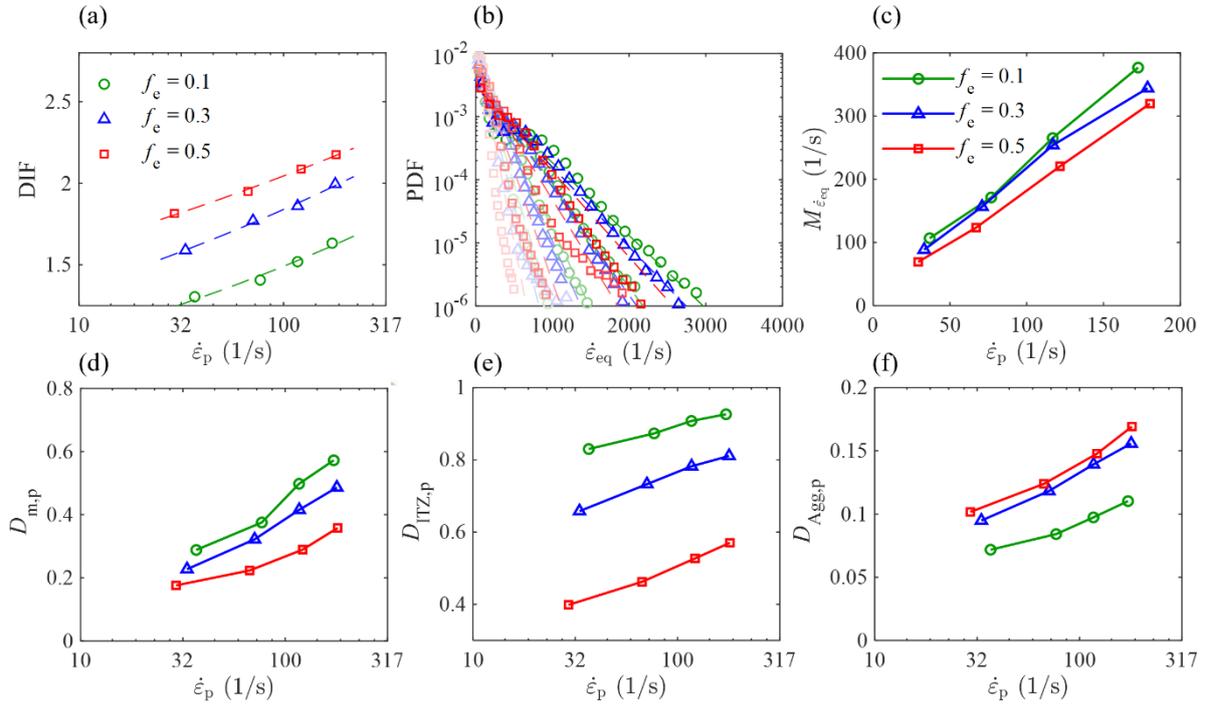

Figure 6: Effects of internal friction, $f_e$, on the same macro- and micro- responses of concrete shown in (a) and (b-f), respectively.

Internal friction coefficients of $f_e = 0.1$, 0.3 and 0.5 are considered for the concrete model. As in Fig. B.2(a) and (b), when the $V_L$ remains the same across different cases, the incident waves are almost identical, while the reflected and transmitted waves dominate the dependence of stress–strain behaviours on $f_e$ under a given $\sigma_{ld,max}$. To highlight the effects of $f_e$ on concrete strength for all dynamic cases, the DIF is redefined, with all values calculated relative to the static strength of concrete with $f_e = 0.1$, as shown in Fig. 6(a), where the $\beta$ values for the slope of DIF-$\dot{\varepsilon}_p$ are measured as 0.148, 0.132, and 0.101 for cases with $f_e = 0.1$, 0.3, and 0.5, respectively. As observed, when a higher $f_e$ leads to higher DIF values, it results in a smaller increase in DIF with increasing $\dot{\varepsilon}_p$, as reflected by the smaller $\beta$ values.

The evolution of the $\dot{\varepsilon}_{eq,avg}$ profile is computed for different $\sigma_{ld,max}$ and $f_e$, as shown in Fig. C.1 of Appendix C. The comparison of Fig. C.1(a-d) with (e-h) or (i-l) shows that the effect of internal friction on this evolution is insignificant and does not become more pronounced or weakened with increasing $\sigma_{ld,max}$. However, these observations cannot be directly linked to the peak stress. Alternatively, this aspect has been further investigated using the semi-plot of PDF and $M_{\dot{\varepsilon}_{eq}}$ in Fig. 6(b) and (c), respectively, both confirming effects of $\sigma_{ld,max}$ and $f_e$. Higher $f_e$ imposes stronger constraints on local deformation, thereby leading to a contraction of the PDF toward lower values with fewer elements reaching high $\dot{\varepsilon}_{eq}$ levels, and correspondingly resulting in a smaller $M_{\dot{\varepsilon}_{eq}}$. Unlike the discussion in Section 3.1, this correlation between $M_{\dot{\varepsilon}_{eq}}$ and $f_e$ is unubiquitous for the absolute values of DIF, providing different corresponding quasi-static strengths at different settings of $f_e$. $f_e$. When an increase in $\sigma_{ld,max}$ induces higher stress acting on the elements, although it results in greater deformation, the constraining effect due to internal friction becomes more pronounced, as evidenced by increased differences in the PDF and $M_{\dot{\varepsilon}_{eq}}$ among various $f_e$. Ultimately, higher $f_e$ results in a less significant increase in $M_{\dot{\varepsilon}_{eq}}$ with increasing $\dot{\varepsilon}_p$, reflecting a smaller slope of DIF-$\dot{\varepsilon}_p$ in Fig. 6(a).

Variations in local damage with $\dot{\varepsilon}_p$ across different $f_e$ are shown in Fig. 6(d-f). With increasing $f_e$, $D_{ITZ,p}$ increases while $D_{Agg,p}$ decreases, reflecting a competitive mechanism between ITZ and aggregate. This means that a higher $f_e$ induces stronger stress concentrations near aggregate surfaces, promoting greater fracture propagation from the ITZ into the aggregate and eventually enhancing the DIF, although a higher $f_e$ can delay fracture initiation and propagation in the mortar and induce a smaller $D_{m,p}$. When local damage increases with increasing $\dot{\varepsilon}_p$, the competition between $D_{ITZ,p}$ and $D_{Agg,p}$ does not become significantly amplified or weakened and therefore cannot explain the effect of internal friction on the slope of the DIF-$\dot{\varepsilon}_p$. This can be attributed to the damage status in mortar. A higher $f_e$ causes a less significant increase in $D_{m,p}$ with increasing $\dot{\varepsilon}_p$. This means that when increasing $\sigma_{ld,max}$ raises contact stresses and promotes greater fracture in the mortar, the constraining effect of internal friction is amplified. Consequently, a higher $f_e$ more strongly inhibits the increase in fracture within the mortar, mapping to a smaller enhancement of the DIF with increasing $\dot{\varepsilon}_p$.

## 3.3. Confining pressure

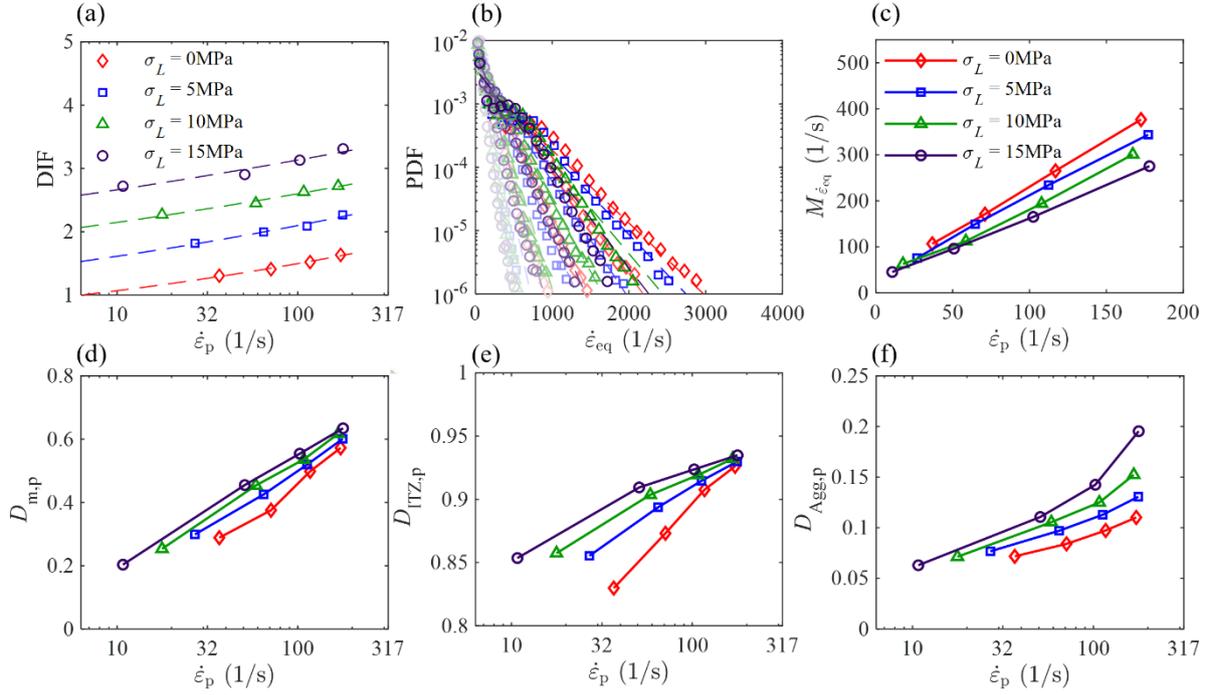

Figure 7: Effects of confining pressure, $\sigma_L$, on the same macro- and micro- responses of concrete shown in (a) and (b-f), respectively.

Here, we consider the confining pressure $\sigma_L = 0, 5, 10$ and 15 MPa to explore the effect of $\sigma_L$. Increasing $\sigma_L$ mainly acts on the reflected and transmitted waves, causing a more ductile strain-stress response of concrete, as shown in Fig. B.3(a-b) of Appendix B. DIF for all cases are calculated as the ratio of the confined dynamic strength to the unconfined static strength of concrete to highlight the effect of confining pressure in Fig. 7(a). The $\beta = 0.147, 0.115, 0.0832$ and 0.0707 are obtained for the slope of DIF-$\dot{\varepsilon}_p$ at $\sigma_L = 0, 5, 10$ and 15 MPa, respectively. Compared with the unconfined case, the confined cases exhibit higher DIF values but show a less pronounced increase in DIF with increasing $\dot{\varepsilon}_p$, as evidenced by the smaller $\beta$ value, similar to findings from experimental studies on confinement [44-47].

The evolution of the $\dot{\varepsilon}_{eq,avg}$ profile for different $\sigma_{ld,max}$ and $\sigma_L$ are shown in Appendix C. As increasing $\sigma_L$ imposes stronger lateral restraint on the element deformation, it shrinks the PDF toward smaller $\sigma_{eq}$ values, with an increasing proportion of elements concentrated in this range, resulting in a smaller $M_{\dot{\varepsilon}_{eq}}$ in Fig. 7(c). With increasing $\sigma_{ld,max}$, the amplified later restraint makes effects of $\sigma_L$ on PDF more pronounced, as also reflected by the increasing difference in $M_{\dot{\varepsilon}_{eq}}$ between different $\sigma_L$. Eventually, higher $\sigma_L$ shows a less significant increase of $M_{\dot{\varepsilon}_{eq}}$ with an increasing $\dot{\varepsilon}_p$, reflecting a smaller slope of DIF-$\dot{\varepsilon}_p$ in Fig. 7(a).

Variations in local damage with $\dot{\varepsilon}_p$ across different $\sigma_L$ are presented in Fig. 7(d-f). As increasing $\sigma_L$ requires higher overall stress to promote the initiation and propagation of fractures due to stronger lateral restraint, damage in all material phases increases; meanwhile, the DIF is also enhanced. When damage increases with increasing $\dot{\varepsilon}_p$, $D_{Agg,p}$ shows more significant effect of $\sigma_L$, in contrast to $D_{ITZ,p}$. This because when increasing $\sigma_{ld,max}$ drives ITZ to gradually reaches a fully fractured state, more fracture propagates from ITZ to aggregate, causing a much higher $D_{Agg,p}$ at higher $\sigma_L$ and retain a higher DIF. Thus, higher $\sigma_L$ shows a more significant increase of $D_{Agg,p}$ with an increasing $\dot{\varepsilon}_p$. $D_{Agg,p}$ cannot directly reflect smaller slope of DIF-$\dot{\varepsilon}_p$ at higher $\sigma_L$. This could be attributed to the mortar. Higher $\sigma_L$ shows a less significant increase of $D_{m,p}$ with increasing $\dot{\varepsilon}_p$, indicating a weakened strain rate effects and a smaller slope of DIF-$\dot{\varepsilon}_p$.

## 3.4. Discussion

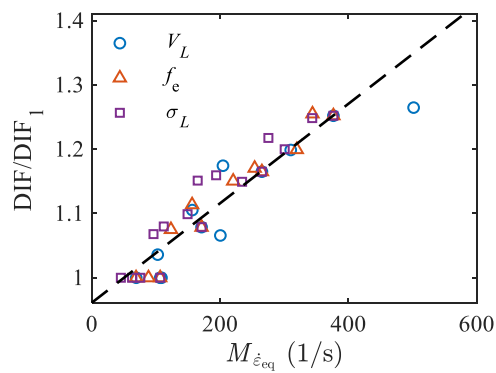

Figure 8: Correlation between $M_{\dot{\varepsilon}_{eq}}$ and normalised DIF.

As discussed in all earlier cases all cases with varying loading ramp rates, $V_L$, internal friction, $f_e$, and confining pressure, $\sigma_L$, there are individual trends between the microscopic statistical index, $M_{\dot{\varepsilon}_{eq}}$, and the overall dynamic strength, DIF. When the distribution of internal strain rate at the peak stress is statistically analysed, variations in $M_{\dot{\varepsilon}_{eq}}$ with $\dot{\varepsilon}_p$ are found to reflect the slope of DIF-$\dot{\varepsilon}_p$ for effects of loading ramp rates, $V_L$, internal friction, $f_e$, and confining pressure, $\sigma_L$. To further confirm this, $M_{\dot{\varepsilon}_{eq}}$ and DIF are collected from different cases to investigate the correlation between both values, as shown in Fig. 8. DIF is normalised by $DIF_1$, the first value of each case, to focus on the relative change respective to the distinct initial values. We found a strong positive correlation between DIF/$DIF_1$ and $M_{\dot{\varepsilon}_{eq}}$ with a correlation coefficient of $R = 0.94$. Thus, the heterogeneity of internal strain rate can be more confidently considered as microscopic indicator for explaining the strain rate effects on the concrete fractures, highlighting the necessity of mesoscale modelling of concrete.

It is noteworthy that the internal strain rate analysed in this study is obtained from mesoscale modelling results that incorporates the rate-independent response of the ITZ due to the limited available data for the material phase [62, 80, 81]. These aspects remain to be implemented to achieve more realistic predictions and to further highlight the role of internal strain rate in the dynamic response of concrete. Moreover, higher strain rate regions (i.e., $\dot{\varepsilon}_p$>200/s) can be further explored as this range can be of interests for many applications in protective structures. The mesoscale model can be extended to focus on sustainable concrete materials by replacing natural aggregates with coral [36], lightweight [66] or recycled aggregates [37], to enrich the understanding of the effects of heterogeneities in material and internal strain rate under dynamic loading.

## 4. Conclusion

In this paper, the dynamic fracture of concrete containing aggregates with realistic shapes is investigated through mesoscale modelling of SHPB tests based on the FEM. The macroscopic dynamic response of concrete has been validated, and we focus on the relationship between the DIF and strain rate with respect to loading ramp rate, internal friction, and confining pressure. Particular attention is given to explaining the DIF related results through microscopic analyses based on the distributions and evolutions of internal strain rate and local damage. The main conclusions are summarised as follows:

- As the loading ramp rate increases, the internal strain rate distribution and the local damage both increase, leading to a higher DIF. With an increase in strain rate of concrete specimen, the damage of both mortar and aggregate exhibits rate-dependent trends similar to internal strain rate distribution. Both show enhanced strain rate effects at higher loading ramp rates, contributing to a more significant increase in the DIF with increasing strain rate.
- With increasing internal friction, although the internal strain rate is more strongly suppressed, the competition between damages of ITZ and aggregate indicates higher induced stress concentration, which promotes fracture propagation from the ITZ into the aggregates, resulting in a higher DIF. Such competition cannot explain the variation in the slope of the DIF-strain rate, as it is mainly attributed to the mortar phase, as evidenced by local damage rather than the internal strain rate. Higher internal friction weakens the strain rate effects on damage of mortar, contributing to a less significant increase in the DIF with increasing strain rate.
- Despite higher confining pressure suppressing the internal strain rate, it can still enhance the DIF, evidenced by increased damage in all material phases. As local damage provides more direct evidence than the internal strain rate, it shows that the less significant increase in the DIF with increasing strain rate at higher confining pressure is mainly attributed to the mortar, which exhibits a weakened strain rate effect on damage under higher confining pressure, although fracture is further promoted in the aggregate than in the mortar at higher strain rates, leading to a more pronounced effect of confining pressure on damage of aggregate.

This mesoscale numerical investigation examines loading and material factors that are difficult or impossible to access experimentally and clarifies their effects on the dynamic response of concrete at both macroscopic and microscopic levels. The analyses of heterogeneities of microscopic information during the dynamic loading provide several insights on explaining the macroscopic observed rate-dependent behaviours. These findings enrich the understanding of strain rate effects in concrete fracture and provide a theoretical basis for improving constitutive relationships of concrete under dynamic loading. Nevertheless, the present work considers only the effects of individual factors, and future studies are required to further develop the mesoscale model to investigate their coupled effects, thereby enabling more accurate predictions of concrete behaviour under complex dynamic conditions.

# Appendix A: Modelling of static fracture of concrete

To simulate concrete fracture under quasi-static conditions, the loading rate is carefully chosen to minimise initial dynamic effects, ensuring that the ratio of total kinetic energy to internal energy remains below 5% in the simulation. Two steel loading plates, with a side length of $L_p = 60$ mm and a thickness of $t_p = 4$ mm, are placed at both ends of the cylindrical concrete specimen, as shown in Fig. A.1, and are meshed using 4 mm hexahedral solid elements. The contact between the loading plate and the specimen is defined as hard contact with a friction coefficient of 0.5. Here, we focus only on the unconfined static cases. The bottom plate is restrained in all directions, while the top plate is allowed to move only in the vertical (loading) direction. The axial stress $\sigma_1$ is applied to the top plate using a velocity-controlled method, in which the velocity of the nodes on the top surface is linearly ramped up to 50 mm/s within 0.001 s and then held constant until the end of the simulation. The step duration is set to 0.02 s.

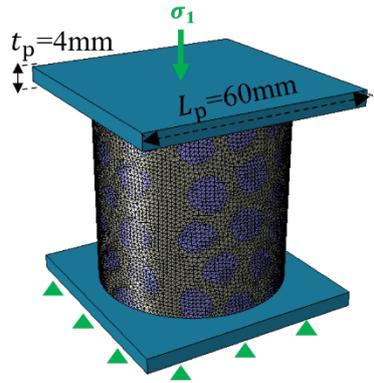

Figure A.1: Schematic diagram of the FE model of quasi-static compressive concrete fracture, where the arrow indicates the direction of the axial loading acting on the plate.

# Appendix B: Strain wave and stress-strain curve

Figs. B.1-B.3(a) show variations in the incident $\varepsilon_{w,i}(t)$, reflected $\varepsilon_{w,r}(t)$, and transmitted $\varepsilon_{w,t}(t)$ strain waves with loading ramp rate $V_L$, internal friction $f_e$, and confining pressure $\sigma_L$, respectively. These are examples when the maximum stress of trapezoidal loading amplitude $\sigma_{ld,max}$ is 90 MPa. The dynamic stress-strain curves are calculated based on Eqs. (5-7) to highlight effects of $V_L$, $f_e$, and $\sigma_L$ under $\sigma_{ld,max}$ = 50, 90, 150, and 250 MPa, as shown in Figs. B.1-B.3(b), where colours from light to dark represent increasing $\sigma_{ld,max}$.

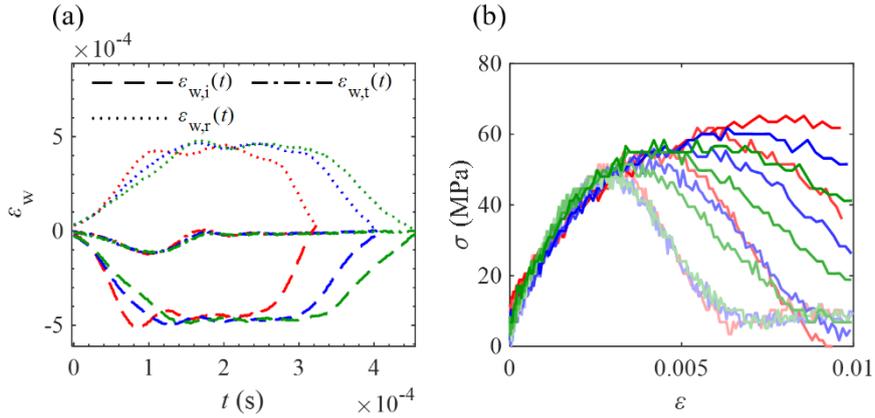

Figure B.1. **Effects of $V_L$ on macroscopic dynamic response of concrete**: (a) Examples of the incident $\varepsilon_{w,i}(t)$, reflected $\varepsilon_{w,r}(t)$, and transmitted $\varepsilon_{w,t}(t)$ waves; and (b) Stress ($\sigma$)-strain ($\varepsilon$) curves for different $\sigma_{ld,max}$ and $V_L$, where cases with $V_L$ = 7692, 10000 and 20000 MPa/s are shown in green, blue, and red, respectively.

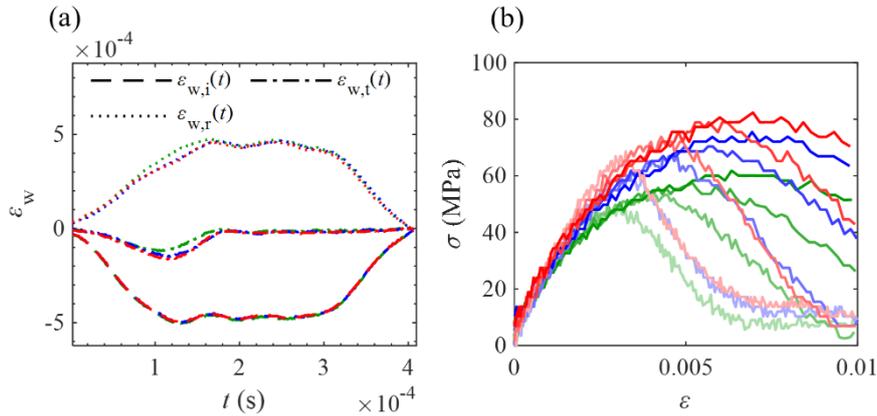

Figure B.2. **Effects of $f_e$ on macroscopic dynamic response of concrete**: (a) Examples of the incident $\varepsilon_{w,i}(t)$, reflected $\varepsilon_{w,r}(t)$, and transmitted $\varepsilon_{w,t}(t)$ waves; and (b) Stress ($\sigma$)-strain ($\varepsilon$) curves for different $\sigma_{ld,max}$ and $f_e$, where cases with internal friction $f_e$ = 0.1, 0.3, and 0.5 are shown in green, blue, and red, respectively.

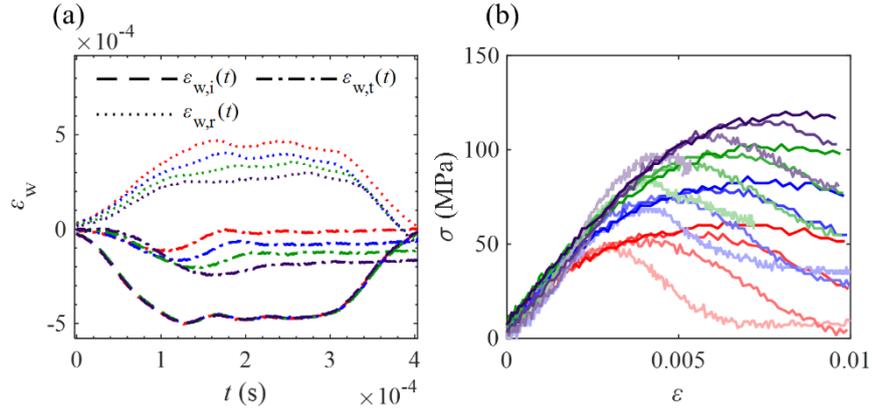

Figure B.3. **Effects of $\sigma_L$ on macroscopic dynamic responses of concrete**: (a) Examples of the incident $\varepsilon_{w,i}(t)$, reflected $\varepsilon_{w,r}(t)$, and transmitted $\varepsilon_{w,t}(t)$ waves; and (b) Stress ($\sigma$)-strain ($\varepsilon$) curves for different $\sigma_{ld,max}$ and $\sigma_L$, where cases with confining pressure $\sigma_L = 0$, 5, 10 and 15 MPa are shown in red, blue, green, and purple, respectively.

# Appendix C: Internal strain rate

Figs. C.1 and C.2 show the time evolution of the profile of the average internal strain-rate distribution, $\dot{\varepsilon}_{eq,avg}$, along the length of the deformed specimen. These are computed using the same method as that applied in Fig. 4 and exhibit similar patterns. Figs. C.1 and C.2 show cases with internal friction values of $f_e = 0.1$, 0.3, and 0.5, and confining pressures $\sigma_L = 0$, 5, 10, and 15 MPa, respectively. when $\sigma_{ld,max} = 50$, 90, 150, and 250 MPa. Figs. C.1 and C.2 are used to support the discussions in Sections 3.2 and 3.3.

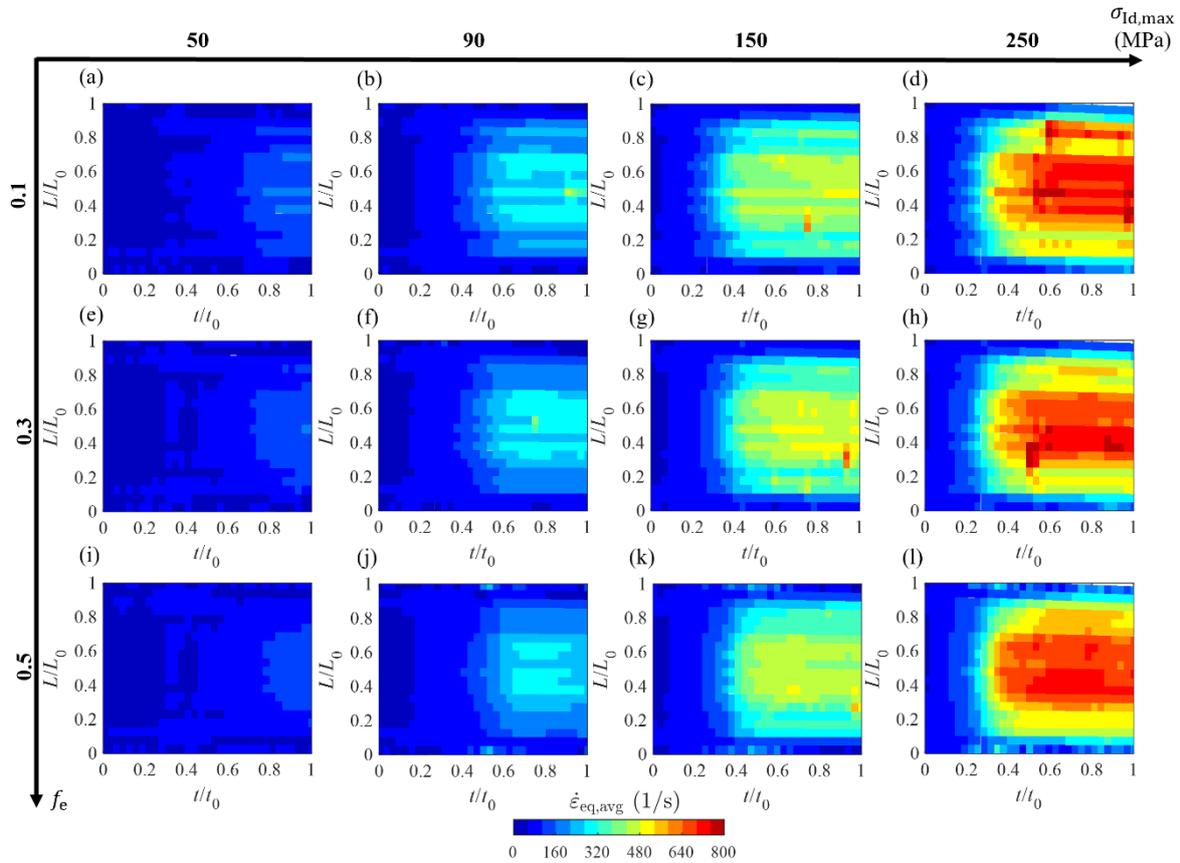

Figure C.1: Evolution of the profile of average internal strain rate, $\dot{\varepsilon}_{eq,avg}$, along the normalised length, $L/L_0$, of the deformed concrete specimen for different $\sigma_{ld,max}$ and $f_e$.

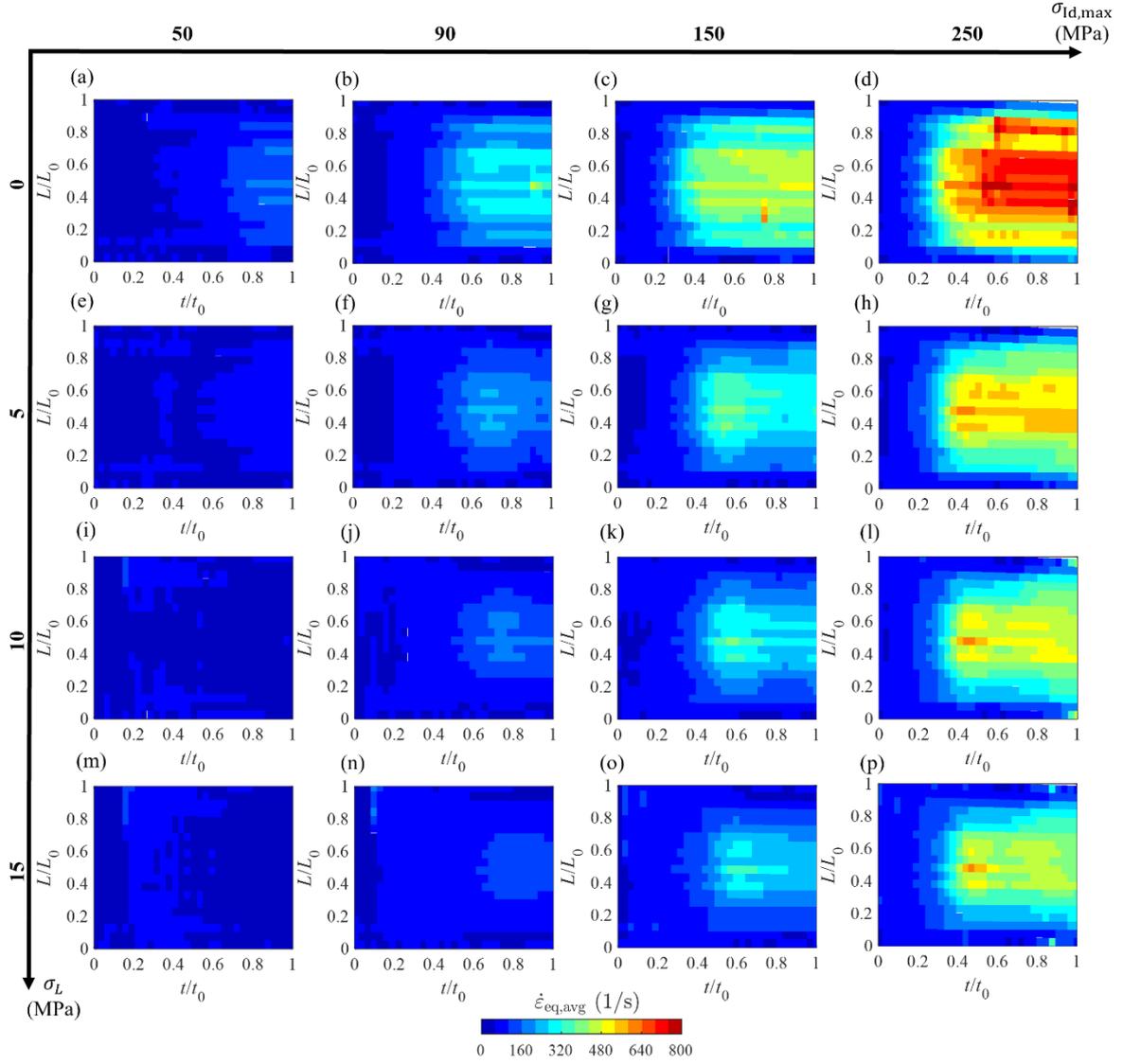

Figure C.2: Evolution of the profile of average internal strain rate $\dot{\varepsilon}_{\text{eq,avg}}$ along the normalised deformed length, $L/L_0$, of the concrete specimen for different $\sigma_{\text{ld,max}}$ and $\sigma_L$.